\documentclass[twocolumn,showpacs,preprintnumbers,amsmath,amssymb]{revtex4}
\usepackage{epsfig}
\usepackage{graphicx}% Include figure files
\usepackage{dcolumn}% Align table columns on decimal point
\usepackage{bm} % bold math
\newcommand{\beq}{\begin{equation}}
\newcommand{\eeq}{\end{equation}}
\newcommand{\beqa}{\begin{eqnarray}}
\newcommand{\eeqa}{\end{eqnarray}}

\def\jpb#1{{ J.\ Phys.\ B} {\bf#1}}

\def\pra#1{{ Phys.\ Rev. A\/} {\bf#1}}

\def\prl#1{{ Phys.\ Rev.\ Lett.} {\bf#1}}

\begin{document}

\title{Multielectron Effects in Sequential Double Ionization with Elliptical Polarization}

\author{Xu Wang}
\email{wangxu@pas.rochester.edu}
\author{J.\ H.\ Eberly}
\affiliation{ Rochester Theory Center and the Department of Physics
\& Astronomy\\
University of Rochester, Rochester, New York 14627}

\date{\today}

\begin{abstract} Multielectron double ionization with elliptically polarized light is examined theoretically using a classical ensemble method going beyond the single-active-electron approximation. A knee structure in the ion signal is found. Our analysis provides an explanation for newly observed SDI phenomena.
\end{abstract}


\maketitle

%===================================================
%\section{Introduction}\label{introduction}
%===================================================

Double ionization studies are revealing and exploiting intriguing new atomic phenomena by use of intense short elliptically polarized pulses \cite{Maharjan-etal, Pfeiffer-etal-1}, and theoretical results are also being announced \cite{Shvetsov-Shilovski-etal, Wang-Eberly09, Wang-Eberly10}. Sequential double ionization (SDI) is  usually described by a single-active-electron (SAE) approximation \cite{Keldysh,PPT,ADK}. It assumes that only one electron is actively involved in the ionization process at any time, and denies other electrons a role other than screening the nucleus. After one electron is ionized, a second electron can become active and subsequently ionized.

An important point that needs to be stressed is that the SAE approximation does not mean no interaction between electrons. Nuclear screening is a consequence of electron-electron (e-e) interaction. Without e-e interaction, nuclear screening is not a valid concept. Therefore what the SAE approximation actually means is a constant interaction between electrons, instead of no interaction.

\begin{figure}[h!]
  \includegraphics[width=6cm]{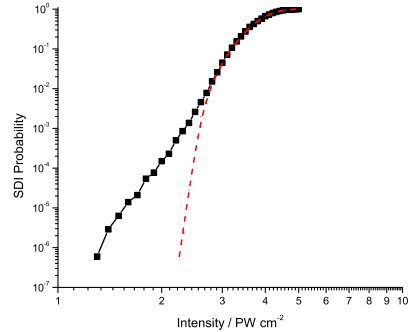}\\
  \caption{\footnotesize SDI probability as a function of intensity. A knee structure can be recognized and identified from multielectron effects. The red dashed curve is inserted by hand to show clearly the knee structure and the intensity range it occupies.}\label{f.SDIknee}
\end{figure}

Using a classical ensemble simulation \cite{ClassicalEnsemble}, we find that the degree of e-e interaction has an intensity dependence. This leads to an unexpected enhancement in the double ionization probability at low intensities. This enhancement appears as a knee structure in the graph of ion count versus intensity, as shown in Fig. \ref{f.SDIknee}. It is very similar to the familiar knee structures induced by recollision correlation \cite{Fittinghoff-etal,Walker-etal,Augst-etal}.

A brief introduction to the classical ensemble method is needed to understand the knee structure shown in Fig. \ref{f.SDIknee}. An ensemble of classically modeled atoms is generated \cite{Abrines, ModelAtom} before turning on a laser pulse. Each atom is composed of a doubly charged fixed ion core at the origin plus two electrons equally able to respond to all forces that are acting. Responses are based on the unbiased solution of the time-dependent Newton equations (TDNE) for the electrons, where unbiased means that speculations about the occurrence of recollision events are never imposed. However, the classical trajectories obtained via the TDNE solutions can be scanned for collision events. To exclude contributions from recollision, a field with ellipticity = 0.5 is used in our simulation and no recollision events are generated.

In the results reported here we took a laser wavelength of 780 nm, and the total energy of the two electrons was set to be -1.6 a.u., which is the sum of the first two ionization potentials of argon. However, the initial positions and momenta of the two electrons are always randomly assigned. Then a laser pulse is turned on and the TDNE solutions are obtained numerically. The size of the ensemble is usually chosen from 1M to 10M to guarantee good statistics. The contribution of the classical ensemble method to understanding strong-field atomic physics has been extensively described \cite{Ho-Eberly06, Haan-etal06,Ho-Eberly07, Haan-etal08,Mauger-etal09}. Recently, in extensions that include elliptical polarization, novel effects have been predicted \cite{Wang-Eberly09, Wang-Eberly10}.

\begin{figure}
  \includegraphics[width=5cm]{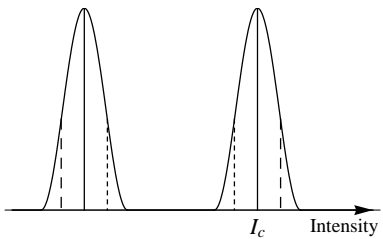}\\
  \caption{\footnotesize Ranges of intensity responsible for ionizations induced by multielectron effects beyond the SAE approximation. The left and right distributions show the ranges of intensity that promote first ionization and second ionization. Two solid lines mark the ionization intensities predicted by an SAE approximation. Due to multielectron effects, the first electron could be ionized with a higher intensity and then the second electron will be ionized with a lower intensity, as indicated by the dotted lines. The opposite ordering of high and low intensities within the upper and lower ranges can also occur, but an inverse relation between first and second intensities is always present.}\label{f.Intensity}
\end{figure}

One of the most important features of the classical ensemble method is that full electron-electron Coulomb interaction is taken into account all the time, from the beginning of the pulse. Therefore multielectron effects beyond the SAE approximation will be included. The degree of interaction between the electrons is not a constant. Instead, it has a wide distribution around the value predicted by the SAE approximation. This uncertainty in e-e interaction strength directly affects ionization of the two electrons. The first electron can be ionized more easily if its trajectory happens to lead to a strong interaction between the two electrons. Then it can take energy from the second electron, and as a result the second electron will be harder to ionize. Or the opposite may happen and the first electron may be more difficult to ionize when interaction with the second electron is weaker. In this case the second electron is easier to ionize due to smaller energy loss. So the ionization difficulty of one electron is inversely related to that of the other electron, as illustrated in Fig. \ref{f.Intensity}.

The SAE approximation provides a critical intensity below which almost no SDI can be generated, which can be thought of as roughly the over-barrier intensity. We denote this intensity as $I_c$ and its position is marked in Fig. \ref{f.Intensity}. With our simulation parameters, we can find from Fig. \ref{f.SDIknee} that $I_c$ takes a value about 2.5 PW/cm$^2$. Above this intensity, there is not much difference between predictions of the classical simulation and the SAE approximation because the second electron is most probably ionized around $I_c$ and additional intensity does not contribute much more to ionization.

However, for laser intensity below $I_c$ where the SAE theory predicts no SDI, active multielectron interactions make a great difference. The first electron can adjust itself to ionize at higher intensity than normal so that the second electron can be ionized with a lower intensity. This is the dotted-line case in Fig. \ref{f.Intensity}. The resulting additional SDI probability manifests itself as the knee structure shown in Fig. \ref{f.SDIknee}. It is interesting to find that this knee structure is induced by a weaker rather than stronger e-e interaction, opposite to what would be normally expected. This SDI knee has not yet been confirmed experimentally.

Further experimental consequences contained in the TDNE results for SDI under elliptical polarization can be identified. We recall the prediction \cite{Wang-Eberly09} that electron pairs ejected in SDI can be expected to show parallel or anti-parallel momenta along the minor elliptical axis direction depending on whether they occur from an in-phase or out-of-phase event, thus providing a 4-peak transverse distribution, and this is evident in the data of Maharjan, et al. \cite{Maharjan-etal}. More detailed examination shows that the ratio of parallel to antiparallel peak heights correlates well with the knee.

\begin{figure} [!b]
  \includegraphics[width=8cm]{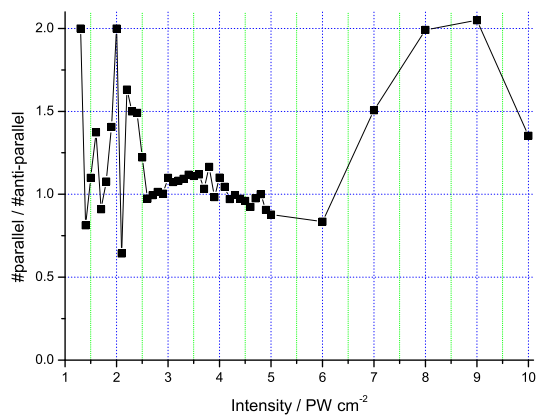}\\
  \caption{\footnotesize Ratio of parallel to antiparallel emitted SDI counts as a function of intensity. Three intensity regions can be clearly identified as described in text.}\label{f.ratio}
\end{figure}

The parallel-antiparallel ratio predicted is shown in Fig. \ref{f.ratio} as a function of intensity. Three quite distinct intensity regions can be recognized. The first region includes intensity up to about 2.5 PW/cm$^2$, within which the ratio oscillates rapidly. The second region is from intensity 2.5 PW/cm$^2$ to about 5.0 PW/cm$^2$, within which the ratio is relatively steady around 1.0. The third region includes intensities greater than 5.0 PW/cm$^2$, within which the ratio oscillates widely but slowly. We find that these three intensity regions coincide with three different knee regimes in Fig. \ref{f.SDIknee}. The first region coincides with the intensity range where the knee structure is evident, up to $I_c$. The second region coincides with the intensity range from $I_c$ to saturation and in the third region SDI is saturated.

\begin{figure}
  \includegraphics[width=6cm]{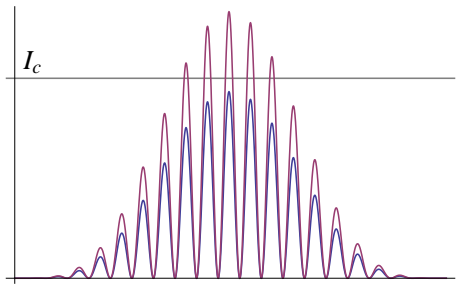}\\
  \caption{\footnotesize Illustration of two pulses with different peak intensities as a function of time. The horizontal line demonstrates the critical intensity $I_c$.}\label{f.pulse}
\end{figure}

It is very attractive to interpret the rapid oscillation of the parallel-antiparallel ratio as a multielectron effect beyond the SAE approximation. We can present a physical mechanism that qualitatively explains how a multielectron effect is needed to cause a rapid oscillation for the low intensities, recalling our analysis of the ionization process in connection with Fig. \ref{f.Intensity}. Two pulses with different peak intensities are shown in Fig. \ref{f.pulse}. One of the pulses peaks above the $I_c$ line and the other below. We focus on the second electron and suppose that ionization of the first electron happened earlier. As explained for Fig. \ref{f.Intensity}, the higher intensity pulse can cause ionization of the second electron around any peak with intensity higher than $I_c$. In this case, parallel and antiparallel SDI peaks happen with almost equal probability. But for the lower intensity pulse, the situation is much different. Only in electron pairs where the first electron received less energy support from the second electron does the second electron retain enough energy to be ionized by the lower pulse. This additional adjustment or selection process makes ionization of the second electron extremely sensitive to intensity. The second electron can virtually only be ionized around the peak intensity. Therefore its ionization is highly localized. This localization of the ionization of the second electron leads to a ratio that can greatly deviate from 1.0. What is more, a slight change in intensity can lead to a significant change in the time difference between the ionizations of the two electrons, and this leads to a rapid oscillation as a function of intensity.

In conclusion, we have shown that with a classical ensemble simulation that allows e-e interaction all the time, multielectron effects beyond the SAE approximation can be identified. A knee structure similar to those induced by recollision processes can appear. Rather than a constant e-e interaction assumed by the SAE approximation, a wide range of possible e-e trajectories means that interaction in the classical ensemble allows a variety of interaction strengths to be experienced. Ironically, the knee structure originates from a weaker e-e interaction than expected from the SAE approximation, instead of a stronger one.

Acknowledgement: this research was supported by DOE Grant No. DE-FG02-05ER15713.

\end{document}